\documentclass[floatfix,aps,showpacs,nofootinbib,amssymb,amsmath]{revtex4}

\usepackage{graphicx}

\usepackage{amsmath}
\usepackage{amsfonts}
\usepackage{amssymb}

\begin{document}

\title{CMB Polarization Systematics, Cosmological Birefringence and the Gravitational Waves Background}

\author{Luca Pagano$^{a}$, Paolo de Bernardis$^{a}$, Grazia De Troia$^{b}$, Giulia Gubitosi$^{a}$, Silvia Masi$^{a}$,
Alessandro Melchiorri$^{a}$, Paolo Natoli$^{b}$, Francesco Piacentini$^{a}$, Gianluca Polenta$^{a,c,d}$}

\affiliation{$^a$ Dipartimento di Fisica and Sezione INFN,
Universit\`a di Roma ``La Sapienza'', Ple Aldo Moro 2, 00185, Rome, Italy}
\affiliation{$^b$ Dipartimento di Fisica and Sezione INFN,
Universit\`a di Roma ``Tor Vergata'', Vle della Ricerca Scientifica 1, 00133, Rome, Italy}
\affiliation{$^c$ASI Science Data Center, ESRIN,  via G. Galilei, 00044, Frascati, Italy}
\affiliation{$^d$INAF-Osservatorio Astronomico di Roma, via di Frascati 33, I-00040 Monte Porzio Catone, Italy}

\date{\today}

\begin{abstract}
Cosmic Microwave Background experiments must achieve very accurate calibration of
their polarization reference frame to avoid biasing the cosmological parameters.
In particular, a wrong or inaccurate calibration might mimic the presence of a gravitational wave background, or
a signal from cosmological birefringence, a phenomenon characteristic of several non-standard, symmetry breaking theories of electrodynamics that allow for \textit{in vacuo} rotation of the polarization direction of the photon. Noteworthly, several authors have claimed
that the BOOMERanG 2003 (B2K) published polarized power spectra of the CMB may hint at cosmological birefringence. Such analyses, however, do not take into account the reported calibration uncertainties of the BOOMERanG focal plane. We develop a formalism
to include this effect and apply it to the BOOMERanG dataset, finding a cosmological rotation angle $\alpha=-4.3^\circ\pm4.1^\circ$. We also investigate the expected performances of future space borne experiment, finding that an overall miscalibration larger then $1^\circ$ for Planck and $0.2\circ$ for EPIC, if not properly taken into account, will produce a bias on the constraints on the cosmological parameters and could misleadingly suggest the presence of a GW background.
\end{abstract}


\maketitle

\section{Introduction}


The latest measurements of the Cosmic Microwave Background
 (hereafter, CMB) anisotropies from satellite, ground based and balloon-borne
experiments such as WMAP (\cite{wmap5}), BOOMERanG-B2K (\cite{boom03}),
ACBAR (\cite{acbar}) and QUAD (\cite{quad}) have revolutionized the field
of cosmology. The precise measurements of
the acoustic oscillations in the angular power spectrum of the
CMB have indeed not only provided a confirmation of the standard
model of structure formation but also constrained key parameters
as the curvature, the energy density in baryons and cold dark matter, and
the shape of the primordial inflationary perturbations.
The new measurements expected by future experiments as Planck will
certainly improve our knowledge of the CMB anisotropies by providing
maps at higher angular resolutions and in a broader frequency spectrum (see e.g. \cite{planck}).

The highest impact for cosmology is however expected not from
temperature but from polarization measurements. The current measurements of
CMB polarization (\cite{wmap5},\cite{boom03},\cite{quad}), while firmly establishing a detection broadly in agreement with
the standard model predictions, lack the precision needed to improve substantially
the cosmological constraints.
Especially on large angular scales, an accurate future measurement of CMB polarization will certainly
shed light on crucial aspects as the epoch and
history of cosmic reionization (\cite{reio}) and the amount of primordial isocurvature
perturbations (\cite{bucher}). More ambitiously,
the CMB polarization is perhaps the most promising tool for
detecting the background of primordial gravitational waves, generated
during inflation \cite{Baumann:2008aq,planck,Bock:2008ww}.

The statistical properties of CMB
linear polarization are indeed fully characterized by two
sets of spin-2 multipole moments with opposite parities. In the standard inflationary paradigm
the primordial magnetic-type modes ($B$ or curl modes) are only produced by
gravity waves (or ``tensor'' metric perturbations \cite{marctwo}), while
density (scalar) perturbations only excite electric-like parity modes ($E$ or
gradient modes) and they do not correlate with each other (see e.g. \cite{szeb}).
A detection of polarization $B$-modes at large angular scales will clearly hint
towards the presence of a gravitational wave background and open a window on
the physics of inflation, at extreme energy scales.

\noindent However, since the seminal paper by Carroll et al. (\cite{carroll})
the possibility of a {\it cosmological birefringence}, or \textit{in vacuo}
rotation of the polarization direction of a photon, has been considered by several authors
(see e.g. \cite{lue,lepora,Xia:2007qs,Kostelecky:2002hh,Finelli:2008jv,Kahniashvili:2008va,Gubitosi:2009eu,2007PhRvD..76l3014C,Balaji:2003sw} and references therein).
Even if exotic, cosmological birefringence is therefore predicted by a wide class of non-standard models and provides
information on symmetry violations and on pseudoscalar fields beyond the standard model.
The main effect of such rotation is to mix the $E$ and $B$ modes.
Considering the usual $C_\ell$ coefficients of the expansions in Legendre polynomials of the corresponding two-point angular
correlation functions (see e.g. \cite{szeb}), the effect of cosmological birefringence is to correlate the spectra by
an angle $\theta$ such that:

\begin{eqnarray}\label{eq:rotPSE}
C_{\ell}'^{TE} &=& C_{\ell}^{TE}\cos (2 \theta) - C_{\ell}^{TB}\sin (2 \theta)  ~\nonumber \\
C_{\ell}'^{TB}&=& C_{\ell}^{TE}\sin (2 \theta) + C_{\ell}^{TB}\cos (2 \theta) ~\nonumber \\
C_{\ell}'^{EE} &=& C_{\ell}^{EE}\cos^2 (2 \theta) + C_{\ell}^{BB}\sin^2 (2 \theta) - C_{\ell}^{EB}\sin (4 \theta) ~\nonumber\\
C_{\ell}'^{BB} &=& C_{\ell}^{BB}\cos^2 (2 \theta) + C_{\ell}^{EE}\sin^2 (2 \theta) + C_{\ell}^{EB}\sin (4 \theta) ~\nonumber \\
C_{\ell}'^{EB}&=& \frac{1}{2}\left(C_{\ell}^{EE}- C_{\ell}^{BB}\right)\sin (4 \theta) + C_{\ell}^{EB}\left(\cos^2 (2 \theta) - \sin^2 (2 \theta)\right)
\end{eqnarray}

\noindent where the new quantities affected by birefringence are denoted with a prime. As we can see, even if one assumes a standard
 scenario where $BB$, $TB$ and $EB$ correlations are zero  at the Last Scattering Surface (LSS), cosmic birefringence introduces a
 new $BB$ component from $EE$. The above equations are exact for a constant rotation, and approximate very well the effects
of a rotation angle varying with time if the time variation is sufficiently slow and the resulting total rotation after
 the propagation of photons (from LSS toward us) is sufficiently small, which is the case for most models of interest
(see \cite{lue,lepora,Xia:2007qs,Kostelecky:2002hh,Kahniashvili:2008va,Gubitosi:2009eu,2007PhRvD..76l3014C,Balaji:2003sw}\footnote{As counterexample see \cite{Finelli:2008jv,Liu:2006uh}}).
The final result is that $B$-modes polarization could therefore not only be sourced by gravitational waves but also from
 a completely different physics.

While both gravity waves and cosmic birefringence would represent sensational discoveries, one has to deal
also with less exciting and definitely more common possibilities as experimental systematics.
Cosmic birefringence could indeed be mostly exactly mimicked by a simple mismatch in the calibration of the principal
axis orientation of the polarimeters.
Let us remind that polarization can be fully described by the four Stokes parameters $I$, $Q$, $U$ and $V$ (see e.g. \cite{Chandra}).
In particular, information about linear polarization\footnote{Circular polarization often neglected since not expected in the
standard scenario} are encoded in the Q and U parameters
that depend on the reference frame. A wrong definition of the principal axis of the polarimeter  which rotates the reference frame of an angle $\theta$ will lead to un-properly defined and mixed $Q$ and $U$, thus resulting in a mixing between $E$ and $B$ modes and producing a change in the polarization power spectra in the same way as described by Equations \ref{eq:rotPSE}.

Cosmic birefringence and how to disentangle it from spurious systematics is therefore a very timely subject of investigation and
has already motivated several recent papers. For example, the possibility of {\it de-rotating} CMB polarization and to reconstruct the rotation angle by measurements of higher-order $TE$, $EE$, $EB$, and $TB$ correlations has been suggested in \cite{marc} and analyzed in detail in
\cite{zalrot}, while in \cite{miller} the effects of beam systematics (which can mimic the rotation of polarization plane)
have been considered.

In this paper we focus on the effects that a rotation of the polarization plane, regardless of its origin, has on the determination
of  cosmological parameters from CMB data when the effect is not properly taken into account. In particular, a miscalibration can
 produce a bias on cosmological parameters as the baryon and cold dark matter densities $\omega_b$ and $\omega_c$, the Hubble constant
$H_0$, the optical depth $\tau$, the inflationary spectral index $n_s$, and a false detection of gravitational waves other than obviously
 be confused with a genuine cosmological birefringence. We also discuss in detail the benefits of inserting a specific parameter describing
 a rotation of systematical origin in the analysis and we employ our formalism on recent data provided by the BOOMERanG-B2K experiment
 as well on simulated datasets of the Planck and EPIC missions. In the case of BOOMERanG-B2K we show that previous marginal detection of cosmological
 birefringence that have been pointed out by \cite{Feng:2006dp}, could be partially explained by a systematic effect.

The paper is organized as follows: in section \ref{sec:syst} we discuss the calibration method adopted for the BOOMERanG-B2K experiment
 and few cases that can induce a systematic rotation of the plane of a polarimeter in future missions;
 in section \ref{sec:method} we set forth our methodology which we
apply in section \ref{sec:B03} to revisit the BOOMERanG-B2K constraints on cosmological birefringence; finally in section
\ref{sec:forecast} we forecast the impact of a systematic rotation on future experiments.


\section{Determining the Polarization Angle: Systematics}\label{sec:syst}

In general, polarimeters for the CMB must detect intensity and
linear polarization. Both can be described through the Stokes I, Q and U parameters (see \cite{Chandra}).
One horned radio-meters and Polarization Sensitive Bolometers (PSB) measure the quantity:
\begin{equation}\label{signal}
V_t = \frac{K}{2} \{I(\hat \gamma_t) + \varepsilon Q(\hat \gamma_t) \cos [2 (\delta+\beta)]
+\varepsilon U(\hat \gamma_t) \sin [2 (\delta+\beta)]\} + n_t
\end{equation}
where $t$ labels time, $K$ is the detector gain,
$\hat \gamma$ the observed sky direction\footnote{More accurately, the signal measured in a
given direction $\gamma$ is given by a convolution between the true sky and an optical response function (or "beam"), which is a tensor
in the case of polarization sensitive detectors. Such detail is irrelevant here, so we simplify the description assuming an infinite resolution
scalar beam}, $n$ is the detector noise, $\delta$ is the
angle between the telescope reference frame and a meridian in the
sky, and $\varepsilon$ and $\beta$ are the polarimeter properties, so defined:
\begin{itemize}
\item $\varepsilon$ is the cross polarization, the ratio between the
minimum and the maximum signal from the polarimeter, when a linearly
polarized source is rotated in front of it;
\item $\beta$ is the angle at which the polarimeter is maximally
sensitive to a vertical linear polarization, in the telescope
reference frame.
\end{itemize}

These parameters must be accurately calibrated.
An incorrect measure of $\varepsilon$ will result in a
fake attenuation or amplification
of the polarization and therefore will induce a bias in the Q and U maps.
An incorrect measure of $\beta$ will
result in leakage from Q to U and from U to Q, thus to a further "mixing" of the
CMB polarization angular power spectra in equations \ref{eq:rotPSE} above: any anaccounted rotation of the polarimeter
will mimic a genuine rotation of the photon polarization plane. There are several techniques to measure the parameters $\beta$ and $\varepsilon$,
depending on the instrument characteristics.\\
In the case of BOOMERanG-B2K, the polarimeters have been
calibrated in the lab prior the flight. The calibration is
obtained using a dedicated apparatus, able to illuminate the full
telescope with a linearly polarized source, modulated by a chopper
wheel. A description of this system is reported in
\cite{Masi_et_al} (see Figure 12 in that paper). The polarization angle of the
source can be rotated as needed to cover the full range $(0-180)^\circ$,
and the angle $\beta$ for each of the detectors has been measured
by means of a best fit procedure. Even if the associated
statistical error was small, there are systematic errors in the
alignment of the calibration source and of the telescope during
the calibration. These, however, affect in the same way all the 16
detectors of the instrument. So if we consider the dataset of the
16 measurements of $\beta$ \cite{piacentiniphd}, the average
value of the difference between nominal and measured orientations
is $\langle \beta \rangle = -0.84^\circ \pm 0.69^\circ$. We assume that
this is due to the above mentioned attitude misalignment. This
value will be used as a prior for the $\beta$ angle in the
following analysis. Instead, the standard deviation of the
differences between nominal and measured orientations is
$\sigma(\beta)=2.75^\circ$ if we include all the 16 detectors, and is
$1.56^\circ$ if we just include only the eight 145 GHz detectors. These
values indicate the effect of rotation errors in the assembly of
the polarimeters, and can be used as an estimate of the precision
of the direction of the main axis for the BOOMERanG-B2K polarimeters.
The axis direction error propagates on the measured power spectra
following eq.1.
So in the following we will mainly consider
the effect of a general rotation of the order of $\langle \beta
\rangle$.\\
In the case of satellite mission such as Planck calibration procedure are even more delicate.
The optical characteristics can change in space due to
thermo-elastic effects, and ground based measurements must be extrapolated
to flight conditions. Alternatively these parameters can be measured
in flight by observation of a strong linearly polarized source.
The most interesting
source from this point of view is the Crab nebula, which
must be observed in advance by some ground based telescope in order to
provide a reliable calibrator.
In the case of Planck HFI, the issue of the polarimeter angle is
complicated even more by the fact that, in the PSB employed, the polarimeter angle
depends on the details of the lithographic
properties of the detector and its coupling to the feeds. For the radio-meters on
board Planck LFI $\epsilon$ is expected to be very close to unity, given to the
excellent isolation provided by the OMT. On the other hand, $\beta$ is subject to similar
potential uncertainties as bolometers and must be calibrated in flight.
For Planck, the uncertainty on the polarimeter calibration
properties are not been disclosed yet. In the following, we assess
what calibration accuracy should be achieved with Planck to yield unbiased
constraints on cosmological parameters.\\


\section{Method}
\label{sec:method}

To understand the impact of a systematic rotation of the CMB linear polarization plane
on the determination of cosmological parameters, we analyze a set of CMB data
with a modified version of the publicly available Markov Chain Monte Carlo
package \texttt{cosmomc} \cite{Lewis:2002ah}. We consider $5$ chains
 with a convergence diagnostics done through the Gelman and Rubin statistic, as now
 common in the literature.
For the CMB data-sets we analyze either the already published BOOMERanG-B2K data,
including the $TB$ and $EB$ spectra from the NA data analysis pipeline (see \cite{boom03}),
either simulated mock data for the ongoing/future Planck\cite{planck} and EPIC\cite{Baumann:2008aq} satellite missions.
For BOOMERanG-B2K we consider two rotation angles ($\beta$ and $\alpha$) in order to take into account
both a systematic and a possible additional angle of true, cosmic birefringence, origin.

For Planck and EPIC we focus on a single rotation angle and we study the impact on the cosmological inferences
drawn by those missions and, in particular, on the detection of a $B$-modes signal from primordial gravity waves.

When simulating the Planck and EPIC future data we assume as fiducial model the maximum likelihood WMAP5 parameters
but with different values of miscalibration angles.
We therefore sample the following eight-dimensional set
of cosmological parameters, adopting flat priors on them: the physical baryon and Cold Dark Matter densities,
$\omega_b=\Omega_bh^2$ and $\omega_c=\Omega_ch^2$, the ratio of the sound
horizon to the angular diameter distance at decoupling, $\theta_s$, the scalar spectral index
$n_s$, the overall normalization of the spectrum $A_s$ at $k=0.05$
Mpc$^{-1}$, the optical depth to reionization $\tau$, the tensor to scalar
ratio of the primordial spectrum $r$ and the systematic rotation angle $\beta$.
We also use a cosmic age tophat prior as $10 Gyr > t_0 >20 Gyr$.
Furthermore, we consider purely adiabatic initial conditions, we impose
flatness and we treat the dark energy component as a cosmological constant.\\
In the case of the BOOMERanG-B2K dataset we include an additional rotation
angle but we fix $\tau=0.09$ consistent with the WMAP5 measurement (\cite{wmap5}).


\section{Analysis of the BOOMERanG-B2K power spectra}
\label{sec:B03}

Here we consider the $TT$, $TE$, $EE$, $BB$, $TB$, $EB$ power spectra
of the Cosmic Microwave Background
obtained by the BOOMERanG-B2K experiment flown in Jan. 2003 (\cite{boom03}).
This dataset has already being analyzed in previous papers. In particular, in
\cite{Feng:2006dp}, \cite{Xia:2007qs} and \cite{Gubitosi:2009eu} a claim for a $\sim 1.2 \sigma$ indication
for cosmic birefringence from this data was presented.
Here we re-analyze the issue by including an additional
 systematic uncertainty $\beta$ on the rotation angle together with the rotation
 angle $\alpha$ assumed as sourced by a true signal of cosmic birefringence.

While both rotation angles affect the CMB spectra as in Eq. 1, there are few
differences. First of all, only photons coming from the last scattering surface at redshift $z\sim1100$
are affected by the birefringence angle $\alpha$. We therefore do not rotate with $\alpha$ the CMB spectra for
$\ell<23$ since the CMB signal from those angular scales is assumed to be sourced from smaller red-shift $z \sim 10$ when the reionization
 process starts (see discussion in \cite{wmap5}). Second, we do not let the
 systematic angle $\beta$ to vary as a free parameter but
 we impose a gaussian prior on it as $\beta =-0.9^\circ \pm 0.7^\circ$ at $1 \sigma$
 motivated by the discussion presented in sec. \ref{sec:syst}.

 Without systematic effects ($\beta=0$) we found that the BOOMERanG-B2K data,
 with a prior on the optical depth $\tau=0.09$, yields a $\sim 1 \sigma$ evidence for cosmological birefringence
 with $\alpha=-5.2^\circ\pm 4.0^\circ$ at $68 \%$ c.l., in agreement with the previous results of \cite{Feng:2006dp,Gubitosi:2009eu,Xia:2007qs}.

 We have then included the possibility of a systematic angle $\beta$.
 Taking into account $\beta$ and the prior on it we instead find $\alpha=-4.3^\circ\pm4.1^\circ$ at $68 \%$ c.l.,
 marginalizing over $\beta$. As we can see, even if the the weak indication for cosmological birefringence is still present,
 the inclusion of $\beta$ reduces the statistical significance at about $1-\sigma$ level.

\begin{figure}[htbp]
\begin{center}
\includegraphics[scale=0.40]{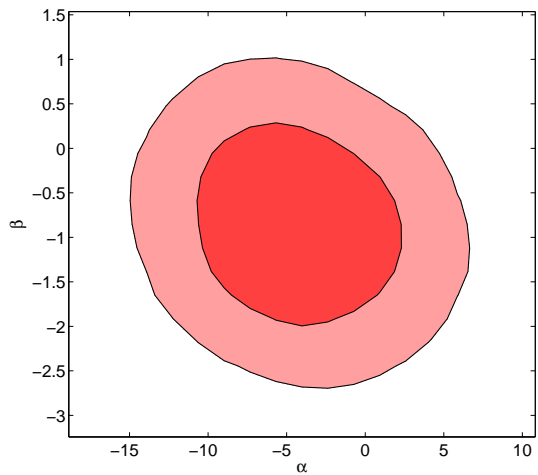}
\caption{Joint 2-dimensional posterior probability contour plot in the $\alpha$-$\beta$ plane, showing the $68 \%$ and $95 \%$
confidence level contours from the BOOMERanG-B2K data.
\label{fig:Boom}}
\end{center}
\end{figure}

It is also interesting to note that from Fig.\ref{fig:Boom}, where we plot the $2$-D likelihood contour plot
from the BOOMERanG-B2K data on the $\alpha$-$\beta$ plane, the probability that the detected rotation is due only to a miscalibration
 ($\alpha=0^\circ$,$\beta=-5.2^\circ$) is suppressed. As we can see in Table \ref{tab:Boom_results} the constraints on the cosmological
parameters from BOOMERanG-B2K data are not much affected by the calibration error, only the upper limit on $r$ is slightly relaxed.

\begin{table}[!htb]
\begin{center}
\begin{tabular}{lll}
Parameter & Not including $\beta$ & Including $\beta$ \\
\hline
$\Omega_bh^2$ & $0.0228^{+0.0068}_{-0.0058}$ & $0.0227^{+0.0069}_{-0.0058} $\\
$\Omega_ch^2$ & $0.123^{+0.063}_{-0.043}$ & $0.122^{+0.063}_{-0.042} $\\
$H_0 $    &     $74^{+25}_{-26}$ & $ 74^{+24}_{-26}$\\
$n_s$ &   $0.88^{+0.21}_{-0.21}$ & $ 0.88^{+0.20}_{-0.21}$\\
$\log[10^{10} A_s]$ &  $3.08^{+0.18}_{-0.16}$ & $ 3.08^{+0.18}_{-0.16}$\\
$r$ &    $<2.6$ & $ <3.4 $ \\
$\alpha$ & $-5.2^{+7.9}_{-8.0}$ & $ -4.3^{+8.2}_{-7.9}$
\end{tabular}
\caption{$95\%$ confidence level on cosmological parameters
obtained by analyzing BOOMERanG-B2K data with and without including the systematic rotation angle $\beta$ in the analysis.}
\label{tab:Boom_results}
\end{center}
\end{table}


\section{Forecast for Planck and EPIC}
\label{sec:forecast}

To forecast the impact of a systematic rotation of the CMB linear polarization plane
on future CMB experiments we create mock datasets
with noise properties consistent respectively with the Planck mission (see \cite{planck})
and the EPIC satellite proposal (see \cite{Baumann:2008aq}).
For the simulated Planck datasets we consider the $100, 143,$ and $217GHz$ HFI
detectors while for EPIC we use the single $150GHz$ channel. The experimental specifications we use are reported in Table~\ref{tab:exp}.\\
We consider for each channel a power noise $w^{-1} = (\theta\sigma)^2$ where $\theta$ is the FWHM of
 the beam assuming Gaussian profile and $\sigma$ is the sensitivity $\Delta T/T$ both from Table~\ref{tab:exp}. We therefore add to each $C_\ell$ spectra a noise spectrum given by:

\begin{equation}
N_\ell = w^{-1}\exp(l(l+1)/l_b^2),
\end{equation}

where $l_b$ is given by $l_b \equiv \sqrt{8\ln2}/\theta$.
The assumed fiducial models are based on the WMAP-5 maximum likelihood parameters {\bf with no primordial
gravitational waves background}. However, we also assume the possibility of a systematic rotation angle $\beta$.
For Planck we consider four fiducial models with
 $\beta= -1^\circ, -2^\circ, -3^\circ, -3.8^\circ$, while for EPIC, that will measure polarization
  with greater accuracy, we consider $\beta=-0.1^\circ, -0.3^\circ, -0.5^\circ, -0.7^\circ$.
 We choose those range of angles since, as we will see below, angles below $\sim 1^\circ$ will not affect
 the cosmic parameters in the case of Planck, while $\sim 4^\circ$ is the current constraint
 on cosmic birefringence (see \cite{wmap5}). In the same way, angles below $\sim 0.1^\circ$ will not affect
 the cosmic parameters in the case of EPIC, while $\sim 1^\circ$ will be presumably the conservative constraint
 on cosmic birefringence at the launch time of this satellite.\\
For each of these datasets we perform two analysis: first we include the systematic
rotation in the montecarlo allowing $\beta$ to vary, then we redo the same analysis without including $\beta$ among
the parameters (i.e. fixing $\beta$ to zero). In the first case we expect to recover the input fiducial model, in
 the second we study the possible bias on the cosmological parameters when the systematic
 miscalibration is not considered. We place particular attention on $r$, since an indication for
 $r>0$ would hint for a detection of gravitational waves,
 while no gravitational waves are assumed in the input fiducial model.\\

\begin{table}[!htb]
\begin{center}
\begin{tabular}{rccc}
Experiment & Channel & FWHM & $\Delta T/T$ \\
\hline
Planck & 70 & 14' & 4.7 \\
$f_{sky}=0.85$& 100 & 10' & 2.5 \\
& 143 & 7.1'& 2.2\\
& 217 & 5' & 4.8 \\
\hline
EPIC & 150 & 5' & 0.81 \\
$f_{sky}=0.85$ & & &\\
\hline
\end{tabular}
\caption{Planck \cite{planck} and EPIC\cite{Baumann:2008aq} experimental specifications.  Channel frequency is given
in GHz, FWHM in arcminutes and noise per pixel in $10^{-6}$ for the Stokes I parameter; the corresponding sensitivities for the Stokes Q and U parameters are related to this by a factor of $\sqrt{2}$.}
\label{tab:exp}
\end{center}
\end{table}

Our results are reported in Table \ref{tab:constraints} where we show the constraints on the standard cosmological
parameters that could be derived from Planck HFI and EPIC respectively with and without considering in data analysis
a possible systematic rotation of the polarization plane.

\begin{figure}[htbp]
\begin{center}
\includegraphics[scale=0.90]{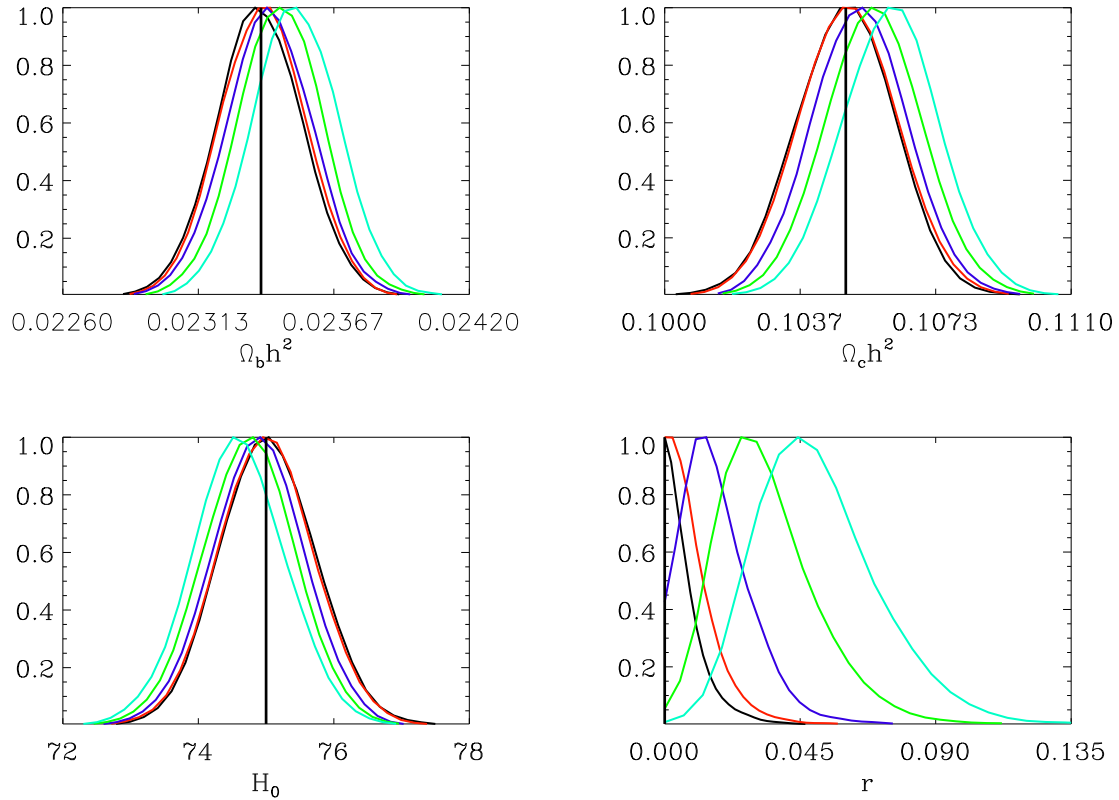}
\caption{Likelihood distribution functions for $\Omega_bh^2$, $\Omega_ch^2$, $H_0$ and $r$
 with (black solid line) and without (other solid lines) including
the rotation angle $\beta$ in the analysis. The noise considered
for the mock data is equivalent to the overall expected sensitivity
($3$ channels) of the Planck mission. The rotation angles considered
in the fiducial model are:
$\beta=-1$ (red), $\beta=-2$ (blue), $\beta=-3$ (green) and $\beta=-3.8$ (cyan)
\label{fig:planck}}
\end{center}
\end{figure}

\begin{figure}[htbp]
\begin{center}
\includegraphics[scale=0.90]{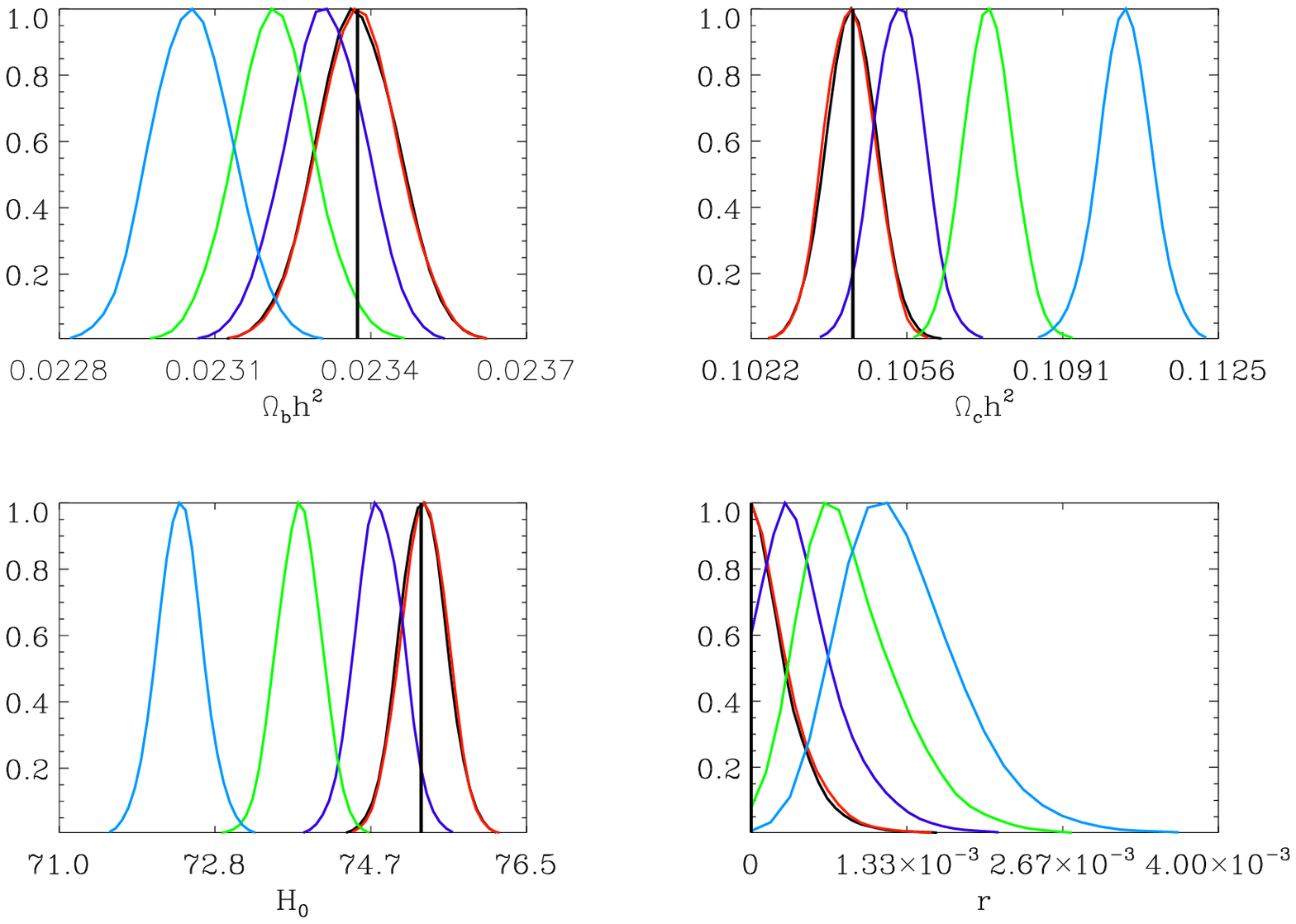}
\caption{Likelihood distribution functions for $\Omega_bh^2$, $\Omega_ch^2$, $H_0$ and $r$
 with (black solid line) and without (other solid lines) including
the rotation angle $\beta$ in the analysis. The noise considered
for the mock data is equivalent to the $150$ GHz Epic channel and
$\Delta \alpha =0.5^\circ$ is assumed in the fiducial model. The rotation angles considered are:
$\beta=-0.1$ (red), $\beta=-0.3$ (blue), $\beta=-0.5$ (green) and $\beta=-0.7$ (cyan)
\label{fig:EPIC}}
\end{center}
\end{figure}

\begin{table}[!htb]
\begin{center}
\begin{tabular}{lccccccc}
\hline
\hline
Parameter & $\omega_bh^2$ & $n_s$ & $\tau$ & $H_0$ & $\omega_ch^2$ & $r$ &$\beta$\\
Fiducial Model&$0.02341$&$0.986$&$0.090$&$75.3$&$0.1046$&$ 0	 $& See below\\
\hline
Planck HFI & & & & & & &\\
\hline
Including $\beta$& & & & & & &\\
$\beta=-1^\circ$&$0.02338^{+0.00032}_{-0.00030}$&$0.9852^{+0.0085}_{-0.0080} $&$0.0900^{+0.011}_{-0.0097}$&$75.0^{+1.3}_{-1.2}$&$ 0.1049^{+0.0024}_{-0.0025}$&$ <0.022 $&$-0.998^{+0.083}_{-0.083} $\\
$\beta=-2^\circ$&$0.02337^{+0.00033}_{-0.00032} $&$ 0.9853^{+0.0083}_{-0.0081}$&$0.090^{+0.010}_{-0.010} $&$75.0^{+1.3}_{-1.3}$&$ 0.1049^{+0.0024}_{-0.0024} $&$<0.021$&$ -2.001^{+0.083}_{-0.080} $\\
$\beta=-3^\circ$&$0.02338^{+0.00033}_{-0.00033} $&$0.9851^{+0.0083}_{-0.0080} $&$0.0895^{+0.011}_{-0.0010} $&$75.0^{+1.3}_{-1.3} $&$0.1049^{+0.0025}_{-0.0025} $&$ <0.022 $&$-3.002^{+0.084}_{-0.087} $\\
$\beta=-3.8^\circ$&$0.02338^{+0.00032}_{-0.00032} $&$0.9852^{+0.0081}_{-0.0085} $&$0.0897^{+0.011}_{-0.0099} $&$ 75.0^{+1.3}_{-1.3} $&$0.1049^{+0.0025}_{-0.0025} $&$ <0.021 $&$-3.801^{+0.084}_{-0.082} $\\
\hline
Not including $\beta$& & & & & & & \\
$\beta=-1^\circ$& $ 0.02339^{+0.00032}_{-0.00032}$& $ 0.9852^{+0.0082}_{-0.0084} $&$0.0898^{+0.011}_{-0.0098}$& $75.0^{+1.3}_{-1.3} $& $ 0.1050^{+0.0025}_{-0.0025}$& $ < 0.026 $& -\\
$\beta=-2^\circ$&$0.02341^{+0.00032}_{-0.00032} $&$ 0.9851^{+0.0084}_{-0.0084}$& $0.0898^{+0.011}_{-0.0097}$& $74.9^{+1.3}_{-1.3} $& $0.1053^{+0.0025}_{-0.0024} $& $<0.039 $&-\\
$\beta=-3^\circ$&$0.02346^{+0.00032}_{-0.00032} $&$0.9856^{+0.0083}_{-0.0083} $&$0.090^{+0.011}_{-0.010} $&$74.7^{+1.3}_{-1.3} $&$0.1057^{+0.0025}_{-0.0024} $&$0.034^{+0.030}_{-0.021} $&-\\
$\beta=-3.8^\circ$&$0.02351^{+0.00032}_{-0.00032} $& $0.9858^{+0.0081}_{-0.0083} $&$0.090^{+0.011}_{-0.010} $&$74.6^{+1.3}_{-1.3} $&$0.1061^{+0.0025}_{-0.0025} $&$0.051^{+0.036}_{-0.027} $&-\\
\hline
EPIC & & & & & & &\\
\hline
Including $\beta$& & & & & & &\\
$\beta=-0.1^\circ$&$0.02341^{+0.00015}_{-0.00015} $&$0.9857^{+0.0045}_{-0.0048}$&$0.0890^{+0.0067}_{-0.0061}$ &$75.26^{+0.51}_{-0.50} $&$0.10444^{+0.0010}_{-0.00098} $&$ <0.00072 $&$-0.0999^{+0.0087}_{-0.0087} $\\
$\beta=-0.3^\circ$&$0.02340^{+0.00015}_{-0.00015}$&$0.9857^{+0.0049}_{-0.0049} $&$0.0888^{+0.0067}_{-0.0066} $&$75.25^{+0.53}_{-0.53} $&$0.1044^{+0.0010}_{-0.0011} $&$<0.00069 $&$-0.2999^{+0.0092}_{-0.0094} $\\
$\beta=-0.5^\circ$&$0.02341^{+0.00015}_{-0.00016} $&$0.9859^{+0.0047}_{-0.0046} $&$0.0891^{+0.0068}_{-0.0065} $&$75.28^{+0.53}_{-0.52} $&$0.1044^{+0.0011}_{-0.0011} $&$<0.00071 $&$-0.5000^{+0.0094}_{-0.0091} $\\
$\beta=-0.7^\circ$&$0.02340^{+0.00015}_{-0.00015} $&$0.9858^{+0.0048}_{-0.0047} $&$0.0890^{+0.0068}_{-0.0067} $&$75.23^{+0.54}_{-0.54} $&$0.1044^{+0.0011}_{-0.0010} $&$<0.00070 $&$-0.6999^{+0.0095}_{-0.0093} $\\
\hline
Not including $\beta$& & & & & & &\\
$\beta=-0.1^\circ$&$0.02340^{+0.00015}_{-0.00015}$&$0.9855^{+0.0046}_{-0.0048}$&$0.0888^{+0.0065}_{-0.0063}$&$ 75.25^{+0.51}_{-0.53}$&$0.1045^{+0.0010}_{-0.0010} $&$<0.00074 $&-\\
$\beta=-0.3^\circ$&$0.02334^{+0.00015}_{-0.00015}$&$0.9831^{+0.0049}_{-0.0049}$&$0.0895^{+0.0068}_{-0.0066}$&$ 74.76^{+0.52}_{-0.53}$&$0.1054^{+0.0011}_{-0.0011}$&$<0.0011 $&-\\
$\beta=-0.5^\circ$&$0.02324^{+0.00015}_{-0.00015}$&$0.9784^{+0.0044}_{-0.0046}$&$0.0908^{+0.0067}_{-0.0063}$&$ 73.82^{+0.49}_{-0.49}$&$0.1074^{+0.0011}_{-0.0010}$&$0.00086^{+0.00078}_{-0.00056}$&-\\
$\beta=-0.7^\circ$&$0.02308^{+0.00015}_{-0.00015}$&$0.9717^{+0.0048}_{-0.0048}$&$0.0925^{+0.0075}_{-0.0067}$&$
72.42^{+0.53}_{-0.51}$&$0.1104^{+0.0011}_{-0.0011}$&$0.00129^{+0.00091}_{-0.00068}$&-\\
\hline
\end{tabular}

\caption{$95\%$ confidence level on cosmological parameters
obtained by analyzing Planck HFI ($100$-$143$-$217$ GHz) and EPIC mock data with different
systematic rotation angles. As we can see the input parameters are always recovered when $\beta$ is included in the analysis.}
\label{tab:constraints}
\end{center}
\end{table}
As we can see in Table \ref{tab:constraints}, the constraints on cosmological parameters based on Planck mission data, when all the $3$ HFI channels are considered , will
be affected by mismatches when $\beta$ is greater than $\sim 1^\circ$ if $\beta$ is not included among parameters in the analysis.
 Instead, as it is shown in Fig.~\ref{fig:planck}, rotation angles of about or lower than $1^\circ$ will have a minimal impact on
the recovered parameters. This result on one hand is reassuring since this is the level
of systematic error expected for this mission but on the other hand it is important to stress that, as will be explained below, there could be an intrinsic rotation of the power spectra due to cosmological birefringence, which could increase (decrease) the total rotation.
However, larger values may bias the result towards lower values for the Hubble constant and baryon density
and larger values for the cold dark matter component. Moreover, a miscalibration
angle of about $\sim 2.5^\circ$ or larger could mimic a detection of B modes at
more than $95 \%$ c.l.. When the rotation angle
is considered in the analysis the input cosmological parameters are correctly recovered. Finally, let us note that the HFI experiment
will be able to measure a systematic angle $\beta$, when considered as an extra parameter, with a precision of
$\sigma(\beta)\sim0.1^\circ$. Since the expected calibration error will be of the order of $\sim 1^\circ$ it is clear that
the use of a markov chain methods such those presented here will definitely be useful in understanding the
experimental calibration.\\
Let us now consider the case for the EPIC satellite proposal.
As we can see
in Table~\ref{tab:constraints} and Fig.~\ref{fig:EPIC}, the sensitivity of this experiment to $\beta$ is
higher respect to Planck, however, rotation angles of about or lower than $0.1^\circ$ will have no effect in
recovering the cosmological parameters. If we do not include $\beta$ in the data analysis, larger values of the systematic angle will have a dramatic effect on all the cosmological parameters, moreover
a miscalibration angle of about $\sim 0.4^\circ$ or larger could mimic a detection of B modes at more than $95 \%$ c.l..
Finally the EPIC experiment will be able to measure a systematic angle $\beta$, when considered as an extra parameter, with a precision of
$\sigma(\beta)\sim0.01^\circ$, an order of magnitude better than Planck HFI.\\

\begin{table}[!htb]
\begin{center}
\begin{tabular}{lccc}
Channel & Calibration angle $\beta$ & \ \ r (including $\beta$) & \ \ r (without $\beta$) \\
\hline
Planck 70GHz &  -1.0 &$<0.27$ & $<0.27 $   \\
Planck 70GHz &  -3.8 &$<0.27 $ & $<0.28 $   \\
\hline
Planck 100GHz  & -1.0 & $<0.082 $ & $<0.087 $  \\
Planck 100GHz &  -3.8 & $<0.082 $ & $<0.13 $  \\
\hline
Planck 143GHz  & -1.0 & $<0.035 $ & $<0.039 $  \\
Planck 143GHz &  -3.8 & $<0.036 $ & $0.052^{+0.045}_{-0.033} $  \\
\hline
Planck 217GHz & -1.0 &  $ <0.077 $ &  $<0.079$ \\
Planck 217GHz &  -3.8 &   $ <0.078 $ & $<0.12 $  \\
\hline
\end{tabular}
\caption{$95\%$ confidence level limits on $r$.}
\label{tab:results}
\end{center}
\end{table}

We analyze in more detail the case of the Planck mission in Table \ref{tab:results}
where we show the effects on $r$ separately for different channels.
For the LFI $70$~GHz channel, the results on $r$ are not dramatically affected even for an unaccounted miscalibration
of $\beta =-3.8^\circ$. The HFI $143$ GHz channel has higher sensitivity and is expected to drive accurate results on cosmological parameters,
even in the case where a set of frequency maps is used to provide a reference foreground cleaned CMB map.
The $143$ GHz channel can, however, be heavily affected by a miscalibration in the polarization angle.
For instance, a value of $\beta = -1^\circ$ will affect the expected bound on $r$ by relaxing it at about
$\sim 10 \%$. A larger miscalibration by $\beta =-3.8^\circ$, if not properly accounted for, will have dramatic implications for this
 channel producing a fake detection of non-zero $B$ modes at more than $2 \sigma$ confidence level. For the other two HFI channels considered
here ($100$ and $217$ GHz) there is no spurious detection of gravitational waves even for a systematic angle as big as $-3.8^\circ$, even if
 the bound on $r$ gets relaxed.
Even a miscalibration $\beta \lesssim 1^\circ$ could, in the case of Planck, yield biased results for $r$ if a genuine cosmological
 rotation ($\alpha$) is present but not accounted for in the analysis. This because an unaccounted rotation can in principle bias $r$
 regardless of its origin (i.e.\ genuine or systematic). As a consequence, it is always necessary to include $\beta$ among the derived
parameters even in the case where the systematic rotation is kept under control.
If a rotation is detected, disentangling its origin (cosmological or not) only based on the data is more troublesome.
If cosmological birefringence is expected to scale with energy, as predicted by some models (see
\cite{Gubitosi:2009eu,Kahniashvili:2008va,Kostelecky:2002hh}) the analysis of multifrequency data can shed light on
the nature of the effect observed, since calibration errors are not normally expected to mimic this behavior. Models that
do not display frequency scaling, on the other hand, cannot be constrained in this way. Another
option is to rely on the $\ell$ scale dependency of the rotation. Since the universe has undergone reionization, a fraction of the
 CMB photons have suffered an effective last scattering at $z \lesssim 20$, thus the magnitude of the accumulated rotations is
expected to be severely suppressed at the lowest multipoles. A systematic rotation, instead, does not display the same behavior
 and hence this fact  could potentially be used to disentangle the two effects. However, when we tried to verify this method on
our Planck simulations we discovered that the number of multipoles that are affected by reionazation is too low for the method to
be useful at disentangling the two contributions. Specifically, we chose not to rotate the multipoles $\ell \leq 23$, to find that
 the $\alpha-\beta$ degeneracy can only be broken this way for amplitudes of order of several degrees, which is totally useless in a
 high precision context. Thus, with the possible exception of exotic models with small scale angular variations,
 models that do not exhibit frequency scaling can only be constrained assuming an accurate enough calibration of the telescope's focal plane.

\section{Conclusions}

CMB polarization will be measured with exquisite precision by future
experiments. A possible and worrying systematic that may affect
this measurement is a miscalibration of the polarization reference frame.
In this paper we have first investigated the impact of this systematic on the
recent claim for a cosmic birefringence signal in the BOOMERanG-B2K
data. We have found that, when the calibration error in the polarization
signal is properly taken into account, the constraint on a cosmic
birefringence angle is $\alpha=-4.3^\circ\pm4.1^\circ$ at $68 \%$ c.l.,
i.e. almost consistent with a statistical fluctuation at one
standard deviation.\\
We have then analyzed the effect of a systematic rotation angle $\beta$
on the determination of cosmological parameters from future CMB surveys,
with a particular attention on the implications for a detection of a GW
background. We have found that for the Planck mission, a miscalibration
 smaller than $\sim 1^\circ$ will have a negligible impact
on the constraints for all parameters, while it could introduce a sizable
bias for larger angles.
For example, systematic error in the calibration of about $\sim 3^\circ$
(i.e. at about $3 \sigma$ from the
expected calibration of $\sim 1^\circ$) would result in a $\sim 3 \sigma$ GW's detection.
We have shown that when including in the usual MCMC analysis the additional parameter $\beta$ to 
consider a possible rotation of the spectra (of systematic or cosmological origin) the bias disappears.
 We have also found that the Planck HFI experiment will be able to constraint
the angle $\beta$ with a precision of the order $\sigma(\beta) \sim 0.1^\circ$ at $95 \%$ and we have investigated
the possibility to disentangle the systematic or cosmological origin of the rotation.
However, the derived sensitivity on $\beta$ is one order of magnitude smaller
than the precision achievable from direct, in-flight, measurements.
It will be therefore mandatory either to calibrate the Planck polarimeters with better than
$1^\circ$ degree accuracy, either to include in future MCMC analysis $\beta$ as an extra parameter.

For the EPIC proposal, a value of $\beta \sim 0.5^\circ$
could heavily affect the cosmological parameters, if not properly
 included in the analysis, leading to a misleading claim of gravity
waves detection at $3 \sigma$ level. On the other hand the
sensitivity on $\beta$ will improve by one order of
magnitude with $\sigma(\beta) \sim0.01^\circ$ at $95 \%$ c.l. in MCMC analysis.

\acknowledgments

This research has been supported by ASI contract I/016/07/0 ``COFIS.''



\newpage

\end{document}